\documentclass[
reprint,
 amsmath,amssymb,
 aps,
 prl, 
 floatfix,
 footinbib,
 sort&compress, numbers, merge
]{revtex4-1}

\usepackage{color}
\usepackage{graphicx}
\usepackage{dcolumn}
\usepackage{bm}
\usepackage{float}
\usepackage{natbib}
\usepackage{hyperref}

\usepackage[usernames,dvipsnames]{xcolor}

\begin{document}

\title{Propulsion and instability of a flexible helical rod rotating in a viscous fluid}

\author{M.K. Jawed$^{1}$}
\author{N.K. Khouri$^{2}$} 
\author{F. Da$^{3}$} 
\author{E. Grinspun$^{3}$} 
\author{P.M. Reis$^{1,2,}$} 
\email{preis@mit.edu}
\affiliation{$^1$Dept.\ of Mechanical Engineering, Massachusetts Institute of Technology, Cambridge, MA 02139, USA\\
$^2$Dept.\ of Civil \& Environmental Engineering, Massachusetts Institute of Technology, Cambridge, MA 02139, USA\\
$^3$Dept.\ of Computer Science, Columbia University, New York, NY 10027, USA
}

\date{\today}%

\begin{abstract}
We combine experiments with simulations to investigate the fluid-structure interaction of a flexible helical rod rotating in a viscous fluid, under low Reynolds number conditions. Our analysis takes into account the coupling between the geometrically nonlinear behavior of the elastic rod with a non-local hydrodynamic model for the fluid loading. We quantify the resulting propulsive force, as well as the  buckling instability of the originally helical filament that occurs above a critical rotation velocity. A scaling analysis is performed to rationalize the onset of this instability. A universal phase diagram is constructed to map out the region of successful propulsion and the corresponding boundary of stability are established. Comparing our results with data for flagellated bacteria suggests that this instability may be exploited in nature for physiological purposes.

\end{abstract}


\pacs{Valid PACS appear here}
\maketitle

Bacteria often rely on the deformation of filamentary helical structures, called flagella, for locomotion~\cite{purcellthe1997, *silverman1977bacterial}. The propulsion arises from a complex fluid-structure interaction (FSI), between the structural flexibility of the flagellum and the viscous forces generated by the flow. This FSI may lead to  
geometrically nonlinear deformations~\cite{turner2000real, *berg2013cell,vogel2012motor}, which in turn can be  exploited for turning~\cite{son2013bacteria}, tumbling ~\cite{macnab1977normal}, bundle formation~\cite{brown2012flagellar} and polymorphic transformations~\cite{calladineconstruction1975, darnton2007force}. 

Resistive force theories (RFT)~\cite{gray1955propulsion,lighthill1976flagellar} are often used to model the role of viscous forces on flexible filaments~\cite{machinwave1958,*takano2003numerical,*takano2003analysis}, 
at low Reynolds number.  These  simplify the viscous loading by introducing local geometry-dependent drag coefficients. More sophisticated descriptions consider non-local hydrodynamic effects, albeit typically assuming that the filament is rigid such that elastic forces are ignored~\cite{rodenborn2013propulsion, spagnoliecomparative2011}. The few studies that have coupled long-range hydrodynamics with elasticity either assume small deflections~\cite{kim2005deformation} or approximate the filament as a network of springs~\cite{flores2005study, olsonmodeling2013}, thereby oversimplifying the mechanics of the problem. One exception is the study of buckling of a straight elastic filament loaded by viscous stresses~\cite{wigginsflexive1998, *coqrotational2008, *qianshape2008, *manghipropulsion2006,
 *wolgemuth2000twirling}. More recently, a systematic computational study has been performed on a discretized model based on Kirchhoff's theory for elastic rods (in the form of a chain of connected spheres), coupled with RFT~\cite{vogel2012motor}. This study was significant in that it was the first, to the best of our knowledge, to report a series of buckling instabilities of the flagellum that arise during locomotion and suggested its relevance to the biological system. Moreover, it addressed the important rotation-translation coupling.
However, recent experiments~\cite{rodenborn2013propulsion, chattopadhyay2009effect, *jung2007rotational} and simulations~\cite{spagnoliecomparative2011} have pointed to the oversimplifying nature of RFT to model propulsion in a quantitatively predictive manner. Therefore, there is a timely need for a  description that fully couples a geometrically nonlinear elastic model of the filament~\cite{kirchhoff1992uber} with long-range hydrodynamic interactions~\cite{lighthill1976flagellar, *lighthillhelical1996,*johnson1980improved}, along with  precision experiments for detailed validation.

\begin{figure}[b!]
\centering
    \includegraphics[width=\columnwidth]{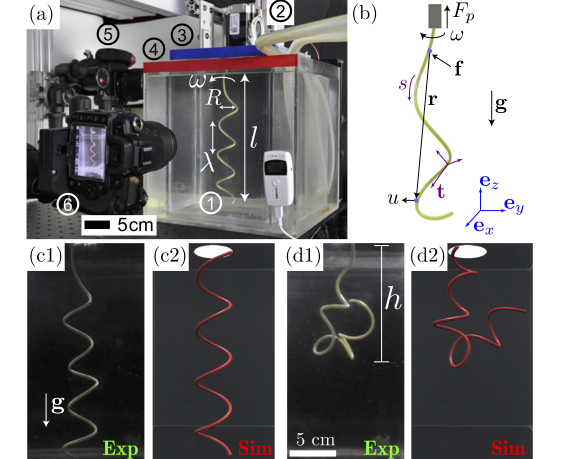}
    \caption{(a) Experimental apparatus: a helical rod~(1), is rotated by a motor~(2), inside a glycerin bath~(3), that is enclosed by an external water tank~(4) for temperature control. Two orthogonal video cameras~(5,6) record the rod.
(b) Schematic diagram of the rod. (c-d) Examples of the deformed rod at $\omega=0.6$ rad/s, from both experiments and simulations:  (c1-c2) helical and (d1-d2) buckled configurations (movie in~\cite{SM}). See text for the properties of rod and fluid.
}
\label{fig:fig1}
\end{figure}

Here, motivated by the locomotion of uniflagellated  bacteria, we perform a combined experimental and numerical investigation of the dynamics of a helical elastic filament rotated in a viscous fluid. 
Our goal is to predictively understand the underlying mechanical instabilities. In our precision model experiments, we reproduce and systematically quantify the dynamics of the filament, as a function of the control and physical parameters of the system. In parallel, we perform numerical simulations that model the elastic rod using the Discrete Elastic Rods (DER) method~\cite{bergou2008discrete, *bergou2010discrete}, coupled to a viscous drag described by Lighthill's slender body theory (LSBT)~\cite{lighthill1976flagellar}. After validating the numerics against  experiments, we  quantify the steady state configurations of the  filament and explore the multi-dimensional phase space of the resulting propulsive force. Existing data on the physical properties of bacterial flagella is sparse~\cite{hoshikawaelastic1983, *hoshikawaelastic1985, *trachtenberg1992rigidity, *flynntheoretical2004, *kim2005deformation}, given the experimental challenges associated with their measurement.  As such, we seek a dimensionless description that encompasses the geometric parameters of natural flagella, with an emphasis on the propulsive force and onset of buckling.
The phase boundary for this instability is mapped out, onto which we locate a number of natural bacterial systems.
These results motivate us to speculate on the potential biological relevance of the mechanical instabilities of rotating flagella.

In Fig.~\ref{fig:fig1}a, we provide a photograph of our apparatus. As a model for flagella, we  cast a series of elastomeric rods with vinylpolysiloxane~\cite{miller2014shapes, *lazarus2013contorting, *lazarus2013continuation} and independently varied each of the geometric parameters (axial length, $l$, or contour length, $L$, helix radius, $R$, pitch, $\lambda$, and cross-sectional radius, $r_0$, or area moment of inertia, $I=\pi r_0^4/4$) and material properties (the Young's modulus, $E$ of the rod was determined by analyzing the shape of a suspended annulus~\cite{adami2013elasto}). The rod was assumed to be incompressible (Poisson's ratio, $\nu \approx 0.5$). During fabrication, a polyvinyl chloride tube was wrapped around a cylindrical object along a helical geometry, and was used as a mold for the rods. The density of the rod was adjusted by adding iron filings (Dowling Magnets) to the polymer, prior to casting. Once cured and demolded, the filament was clamped at one end, immersed in a bath of glycerin (20$\times$20$\times$30$\,\mathrm{cm^3}$), and rotated using a stepper motor~\cite{SM}. Using digital imaging, we reconstructed the deformed configurations of the filament and quantified its dynamics. To ensure constant and reproducible values for the fluid viscosity, the glycerin bath was inserted within an external water tank to accurately control the temperature within $\pm0.5\mathrm{^\circ C}$ (Brinkmann Lauda RC6). 
By tuning the temperature, 7.6$\le\theta[\mathrm{^\circ C}]\le$ 32.4, we varied the viscosity of glycerin in the range 0.50 $\le \mu[\mathrm{Pa\cdot s}] \le$4.45 ($\pm0.05\,\mathrm{Pa\cdot s}$). The density of glycerin is $\rho_m = 1.24~\mathrm{g/cm^3}$, and despite our best effort for density matching, our rods had a slightly higher value ($\lesssim5\%$) than glycerin, which is however included in the numerics.
 
For our simulations, we combined DER~\cite{bergou2008discrete, *bergou2010discrete}, a robust and efficient computational tool for the mechanics of rods, 
 and  LSBT~\cite{lighthill1976flagellar}, a viscous force model that accounts for non-local hydrodynamics. Both DER and LSBT were independently validated against precision experiments in Refs.~\cite{jawed2014coiling} and~\cite{rodenborn2013propulsion}, respectively.
The helical rod is described by its centerline, parameterized by the arc-length, $s$ (Fig.~\ref{fig:fig1}b). For the fluid loading, LSBT is used to relate~\cite{lighthill1976flagellar} the local velocity, $\mathbf{u}(s)$, and the force per unit length, $\mathbf{f}(s)$, at each point on the rod centerline:
\begin{equation}
\mathbf{u}(s) = \frac {\mathbf{f}_{\perp} (s) } { 4 \pi \mu } +
\int_{|\mathbf{r}(s', s) | > \delta } \mathbf{f} (s') \cdot \mathbb{J} (\mathbf{r}) \mathrm{d}s\rq{},
\label{eq:LighthillSBT}
\end{equation}
where $\mathbf{f}_{\perp} (s) = \mathbf{f}(s) \cdot \left( \mathbb{I} - \mathbf{t}(s)  \otimes \mathbf{t}(s) \right)$ is the component of $\mathbf f$ in the plane perpendicular to the tangent, $\mathbf{t}(s)$, $\mathbf{r}(s^\prime, s)$ is the position vector from $s^\prime$ to $s$, $\delta = \frac { r_0 \sqrt e} {2}$ is the natural cutoff length, and $\mathbb J (\mathbf r) = \frac {1} {8 \pi \mu} \left( \frac {\mathbb I} {| \mathbf r|} + \frac {\mathbf r \otimes \mathbf r}  {| \mathbf r|^3 } \right)$ is the Oseen tensor. 
Eq.~(\ref{eq:LighthillSBT}) is then discretized and cast into a $3N$ sized linear system of the form $\mathbf{U} = \mathbf{A F}$, where $N$ is the number of nodes of the discretized rod~\cite{SM}. At each time step in DER, the viscous forces, $\mathbf F$, are evaluated from the velocities, $\mathbf U$, and the matrix $\mathbf A$ that only depends  on the geometric configuration of the rod. To advance in time, we apply this external force together with elastic forces, update the rod configuration  and iterate. Self-contact, possible only after buckling, is neglected throughout, although this does not compromise the agreement with experiments.

\begin{figure}[b!]
\centering
    \includegraphics[width=\columnwidth]{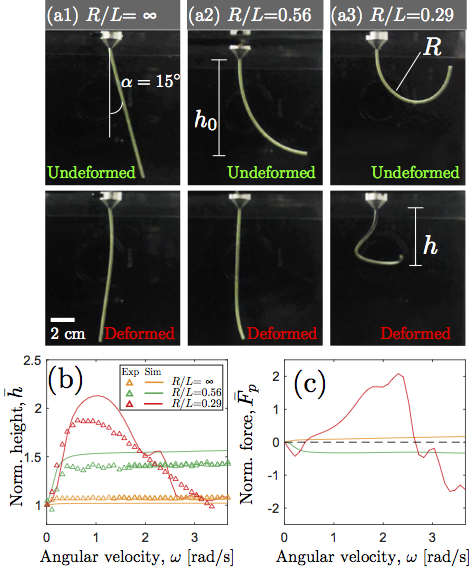}
    \caption{(a) Experimental images of (a1) straight rod, (a2) $R/L=0.56$, and (a3) $R/L=0.29$ in their undeformed (top row) and deformed state (bottom row) at $\omega=3.14$ rad/s. 
(b) Normalized suspended height, $\bar h$, versus angular velocity, $\omega$, for experiments and simulations.
(c) Simulation data of normalized propulsive force, $\bar F_p$, versus $\omega$.
}
    \label{fig:fig2}
\end{figure}

We first establish a connection with existing literature for a naturally straight filament rotating in a viscous fluid~\cite{coqrotational2008,*qianshape2008} and then consider naturally curved rods.
In Fig.~\ref{fig:fig2}a, we present experimental photographs of undeformed (top) and deformed (bottom) configurations, for three representative cases of decreasing the natural radius of curvature of the rod, $R$, while fixing its contour length at $L=12.00 \pm 0.05\,\mathrm{cm}$. These three cases are: i) straight rod (clamped at an angle of $\alpha = 15^\circ$, for consistency with Ref.~\cite{coqrotational2008}), ii) moderately curved rod ($R/L=0.56$, $\alpha=0$) and iii) highly curved rod ($R/L=0.29$, $\alpha=0$). All other parameters for this part of the study were kept fixed: $r_0 = 1.58 \pm 0.02~\mathrm{mm}$, rod density, $\rho_r =1.306 \pm 0.002~\mathrm{g/cm^3}$, $E=1255\pm 49~\mathrm{kPa}$, and $\mu = 1.32 \pm 0.05~\mathrm{Pa\cdot s}$, which ensured a Reynolds number $< 10^{-1}$. The resulting configurations (after initial transients) for these three cases are found to vary dramatically with $R/L$. 

The role of natural curvature is quantified further in Fig.~\ref{fig:fig2}b, where we plot the steady state suspended height (vertical distance from the clamp to the bottom of the rod, $h$) normalized by the height in the  non-rotating case, \emph{i.e. $\bar h=h/h_0$}, as a function of the imposed angular velocity, $\omega$. Excellent quantitative agreement is found between experiments and simulations, with no fitting parameters. Given that the value of the propulsion force at the clamp is too low to be measured experimentally, we extract it from the simulations at each time step as  $F_p = - \int_0^L (\mathbf{f}\cdot\mathbf{e}_z) \mathrm{ds}$, where $\mathbf f$ is obtained from Eq.~(\ref{eq:LighthillSBT}). In Fig.~\ref{fig:fig2}c, we normalize the propulsive force, $\bar{F}_p=F_p L^2/(EI)$, by the characteristic bending force in the rod and plot $\bar{F}_p$ versus $\omega$.  Qualitative and quantitative differences are observed between the three cases: straight, moderately curved, and highly curved rods. The first two undergo a shape transition at $\omega \approx 0.2\,$ rad/s. However, the propulsion force of the straight rod is always positive, whereas it is negative for the moderately curved rod. By contrast, the highly curved rod exhibits a non-monotonic behavior: $\bar{h}$ first increases to reach a maximum value at $\omega \approx 1.0$ rad/s, where buckling  occurs, and eventually $\bar{h} \approx 1$. Since the propulsion depends on the deformed configuration, the resulting $F_p$ vs. $\omega$ relation is markedly different from the previous two cases; $F_p$ first changes sign from negative to positive at $\omega \approx 0.3\,$ rad/s,  reaches a maximum at $\omega \approx 2.4\,$ rad/s and then changes sign again at $\omega \approx 2.6\,$rad/s. The coupled effect of curvature, flexibility and fluid forces can thus produce nontrivial behavior in both geometry and propulsion.

Reassured by the quantitative agreement between  numerics and experiments, we turn to the dynamics of helical filaments as macroscopic analogues of bacterial flagella~\cite{fujii2008polar, spagnoliecomparative2011}. These rods were rotated in the glycerin bath ($\mu = 1.6 \pm 0.05~\mathrm{Pa\cdot s}$) at angular velocities in the range $0<\omega\,[\mathrm{rpm}]\leq8\,$. For now, we focus on a case with: $E=1255\pm 49~\mathrm{kPa}$, $\rho_r = 1.273 \pm 0.022\,\mathrm{g/cm^3}$, $l=20 \pm 0.5\,\mathrm{cm}$, $\lambda=5 \pm 0.5\,\mathrm{cm}$, $R=1.59 \pm 0.1\,\mathrm{cm}$, and $r_0 = 1.58 \pm 0.02\,\mathrm{mm}$.
Under these conditions, the Reynolds number always remains smaller than $10^{-2}$ and the Stokes flow assumption is appropriate throughout.
\begin{figure}[t!]
    \includegraphics[width=\columnwidth]{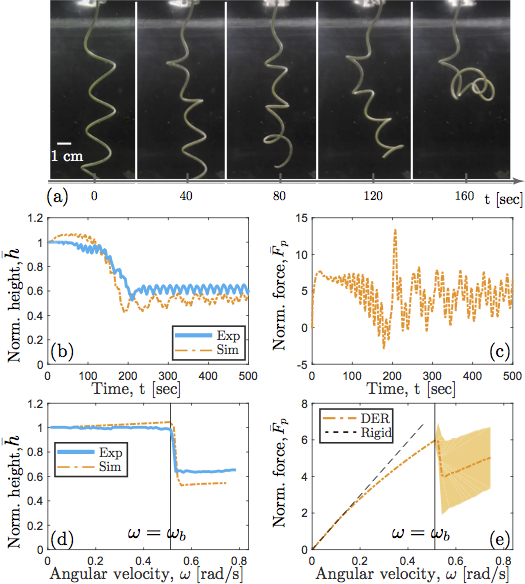}
    \caption{
(a) Sequence of experimental images at $\omega = 0.6$ rad/s. Material properties are provided in the text (movie in ~\cite{SM}).
Time series of (b) normalized height, $\bar h (t)$,
 and (c) normalized propulsive force, $\bar F_p (t)$.
(d) Normalized height, $\bar h$, in steady state ($t>360$ s) versus $\omega$.
(e) Normalized propulsive force, $\bar F_p$, in steady state versus $\omega$, with the shaded region representing the standard deviation.}
\label{fig:fig3}
\end{figure}
In Fig.~\ref{fig:fig3}a, we present a  sequence of experimental photographs for our representative helical rod rotated at $\omega = 0.6\,$rad/s, starting from rest. The corresponding time series for the normalized suspended height, $\bar{h}$, is plotted in Fig.~\ref{fig:fig3}b for both experiments (solid line) and simulations (dashed line), with good agreement between the two.
Any mismatch arises primarily from self-contact that is neglected in the simulations.
Initially ($t\lesssim 100\,$s), $\bar{h}\sim 1$ but the  configuration eventually becomes increasingly distorted due to the appearance of regions of chiral inversion, even if the axis of the helix remains vertical. At later times, the rod bundles and the suspended height reaches an approximate steady state, with $\bar{h} \sim 0.6$. The time series of the normalized propulsive force, $\bar{F}_p = F_p l^2 / (EI)$, calculated from the simulations, is plotted in Fig.~\ref{fig:fig3}c. 
Concurrently with the drop in $\bar h$ at $t \approx 150\,$s, $\bar{F}_p$ becomes increasingly unsteady, which we will show arises through a buckling instability.


\begin{figure*}[t!]
    \includegraphics[width=1.0\textwidth]{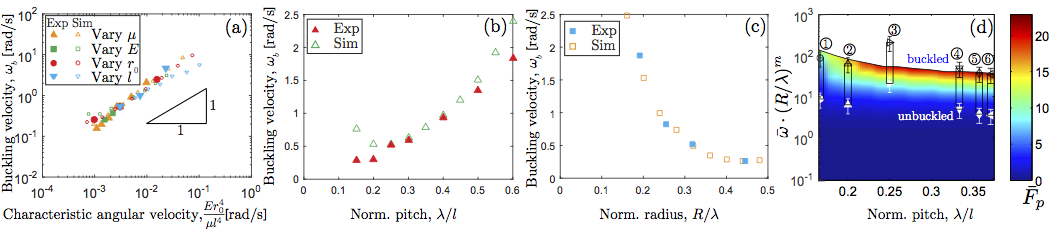}
    \caption{
Buckling velocity, $\omega_b$, versus (a) characteristic angular velocity, $E r_0^4 / \mu l^4$, (b) normalized pitch, $\lambda/l$, and (c) normalized helix radius, $R/\lambda$. In (a), one of the four parameters $\{ E, r_0, \mu, l\}$ was varied keeping the others fixed. In (b) and (c), the helix pitch and radius, respectively, were varied while fixing all the other parameters. 
(d) Dependence of propulsive force (color bar) on both the normalized pitch, $\lambda / l$, and $\bar \omega (R/\lambda)^{m}$, with $m=1.96\pm 0.05$, obtained at $R/\lambda = 0.2$ (see~\cite{SM}).
The symbols correspond to
\textcircled{1} {\em Caulobacter crescentus} (Wild) ($\circ$),
\textcircled{2} {\em Rhizobium lupini} (Curly) ($\triangle$),
\textcircled{3} {\em Salmonella} (Wild) ($\rhd$),
\textcircled{4} {\em Rhizobium lupini} (Semicoiled) ($\bigtriangledown$),
\textcircled{5} {\em Escherichia coli} ($\Diamond$),
\textcircled{6} {\em Vibrio alginolyticus} ($\lhd$)~\cite{rodenborn2013propulsion, chattopadhyay2009effect}.
Filled (and open) symbols correspond to the lower (and upper) bound estimates $EI=10^{-23}~\mathrm{Nm^2}$ (and $EI=10^{-24}~\mathrm{Nm^2}$). The errorbars represent the range in angular velocity, $\omega/(2\pi) = 500\pm 200$ Hz.}
    \label{fig:fig4}
\end{figure*}

In Fig.~\ref{fig:fig3}d, we plot the late time average of $\bar{h}$ over $740$s past the initial transients ($t>360\,$s) versus $\omega$. We find that $\bar{h}\sim 1$ up to $\omega_b=0.51\,$rad/s, after which it sharply drops. Hereafter, we shall refer to $\omega_b$ as the \emph{critical buckling velocity}, above which fluid loading arising due to the rotation causes the helical filament to buckle. The corresponding $\bar{F}_p$ is plotted in Fig.~\ref{fig:fig3}e as a function of $\omega$ and we find that it increases monotonically up to $\omega_b$. Note that a rigid helix would yield a linear dependence between $\bar{F}_p$ and $\omega$~\cite{rodenborn2013propulsion} (dashed line in Fig.~\ref{fig:fig3}e). For $\omega<\omega_b$, flexibility of the helical filament leads to a sub-linear $\bar{F}_p$, when compared to the rigid case. For $\omega>\omega_b$, the average value of $\bar{F}_p$ drops sharply, albeit with significant fluctuations (the shaded region in Fig.~\ref{fig:fig3}e correspond to the standard deviation of the averaged force value). Similar to $F_p$, we can also measure the input torque, $T_p$, necessary to sustain rotation, and this allows us to quantify the propulsive  efficiency, which we define as $\eta = F_p l / T_p$. The efficiency remains almost constant as a function of angular velocity for $\omega < \omega_b$, but drops to a lower value upon buckling~\cite{SM}. Also, see Ref.~\cite{vogel2012motor} for a characterization of propulsion as a function of input torque.

We proceed to rationalize the dependence of the buckling velocity on the physical parameters of our system. 
The viscous force acting on the helical filament scales as $F_v \sim \mu \omega l^2$. Regarding the helix as an effective beam allows us to estimate its critical buckling load as $F_c \sim EI/l^2$. Instability is expected to occur when $F_v \approx F_c$, which yields
%
$\omega_b = \bar{\omega}_b E I / (\mu l^4)$, 
%
where $\bar \omega_b = \hat{\bar{\omega}} (\lambda/l, R/\lambda)$ is a dimensionless function that depends on the geometry of the helix alone. 
To systematically investigate the dependence of $\omega_b$ on the various parameters, we start with the values for the rod used in Figs.~\ref{fig:fig3} and \ref{fig:fig4}a, and plot  $\omega_b$ versus $Er_0^4/ (\mu l^4)$, for a given geometry ($\lambda/l=4$ and $R/\lambda=0.32$). In both experiments (filled symbols) and simulations (open symbols), each one of the four parameters $\{ E, r_0, \mu, l\}$ is independently varied, while fixing the other three. We find that the data collapse onto a straight line, thereby supporting our scaling analysis. 
To characterize the effect of the geometry $\{\lambda/l, R/\lambda\}$, we independently vary the helix radius, $R$, and pitch, $\lambda$, while fixing $E$, $\mu$, $r_0$ and $l$. In Fig.~\ref{fig:fig4}b,  we plot $\omega_b$ as a function of the normalized pitch, $\lambda/l$ at fixed $R = 1.59\,$cm. The dependence of $\omega_b$ on the normalized radius, $R/\lambda$, at $\lambda=5\,$cm is shown in Fig.~\ref{fig:fig4}c. 
From both of these plots, we conclude that $\omega_b$ varies  nonlinearly with  $\lambda/l$ and $R/\lambda$.
These parameters must be taken into account when mapping the results from our model system to a regime that is relevant to bacterial locomotion, which is address next. 

Finally, 
we  take advantage of the efficiency of our algorithm to provide a biologically relevant description of $\bar{\omega}_b$. We use the parameters of the rod used in Fig.~\ref{fig:fig3}, except that the fluid and flagellum are assumed to be density matched and the axial length is increased to  $l=0.4\,$m, so that $r_0 \ll \{\lambda, R, l\}$. Supported by the data (see~\cite{SM}), we approximate $\bar \omega_b = \hat M (R/\lambda) \, \hat N (\lambda / l)$ by $\hat M (R/\lambda) = (R / \lambda) ^{-m}$ with $m = 1.96 \pm 0.05$ (see ~\cite{SM} for details). Using this result, in Fig.~\ref{fig:fig4}d, we construct a phase diagram for the propulsive force, versus both $\bar \omega \cdot (R / \lambda)^m$ and $\lambda / l $, where $\bar \omega = \omega \mu l^4 / (EI)$ is the normalized angular velocity. 
In Fig.~\ref{fig:fig4}d, we also superpose the  parameter values corresponding to bacterial flagella of specific organisms (see caption), which are estimated by taking the characteristic orders of magnitude values for $\mu = 10^{-3}\,\mathrm{Pa.s}$, $EI = 10^{-23}\,\mathrm{N m^2}$~\cite{takano2003analysis} (this estimate ranges from $10^{-24}$~\cite{fujime1972flexural} to  $10^{-22}$~\cite{hoshikawaelastic1985}), and $\omega = 10^2 - 10^3~\mathrm{Hz}$~\cite{chattopadhyay2009effect, magariyama1995simultaneous, son2013bacteria}. Moreover,  the geometric parameters $\{\lambda/l, R/\lambda\}$ for some common bacteria (see caption of Fig.~\ref{fig:fig4}) 
are taken from Refs.~\cite{rodenborn2013propulsion, chattopadhyay2009effect}. The data suggests that  natural flagella rotate at a rate approximately within one order of magnitude of $\omega_b$, where we have taken into account the estimated range of $\omega$ (errorbars) and the known uncertainty in $EI$ (rectangles). Note that for simplicity and generality, we ignored the role of the cell body, and focused on a single helical filament, even though a number of the bacteria considered here are multi-flagellated.

Our results raise the hypothesis that the flexibility of flagella imposes an upper bound on propulsive force through $\omega_b$, above which buckling occurs. 
Moreover, in addition to the localized bucking that can happen at the hook of the flagellum~\cite{son2013bacteria}, the reconfigurations that arise in the post-buckling regime of the flagellum suggest the possibility of a novel functional turning mechanism. This remains an open question, however, given the current uncertainty on the known properties of flagella. As such, our investigation calls for additional experimental work to more precisely measure the  properties of natural bacterial flagella, and more accurately image their dynamics.

\begin{acknowledgments} We thank Roman Stocker for enlightening discussions and we are grateful to the National Science Foundation (CMMI-1129894) for financial support.
\end{acknowledgments}

\bibliographystyle{apsrev4-1}
\bibliography{flagella}

\end{document}